\documentclass[aps,pra,twocolumn,superscriptaddress,showpacs]{revtex4}

\usepackage{amsmath}
\usepackage{amsfonts}
\usepackage{amssymb}
\usepackage{graphics}
\usepackage{graphicx}

\begin{document}


\title{The conservation of orbital angular momentum and the two-photon detection amplitude in spontaneous parametric down-conversion }

\author{Sheng Feng} 
\email{sfeng@ece.northwestern.edu}
\affiliation{Center for Photonic Communication and Computing, EECS 
Department, Northwestern University, Evanston, IL 60208-3118, U.S.A.}
\author{Chao-Hsiang Chen}
\affiliation{Department of Physics and Astronomy, Northwestern 
University, Evanston, IL 60208-3112, U.S.A.}
\author{Geraldo A. Barbosa}
\affiliation{Center for Photonic Communication and Computing, EECS 
Department, Northwestern University, Evanston, IL 60208-3118, U.S.A.}
\author {Prem Kumar}
\affiliation{Center for Photonic Communication and Computing, EECS 
Department, Northwestern University, Evanston, IL 60208-3118, U.S.A.}
\affiliation{Department of Physics and Astronomy, Northwestern 
University, Evanston, IL 60208-3112, U.S.A.}




\date{\today}

\begin{abstract}
We study the two-photon detection amplitude of the down-converted
beams in spontaneous parametric down-conversion when the physical
variable of orbital angular momentum is involved, taking into account
both conservation and non-conservation of angular momentum. Agreeing
with experimental observations, our theoretical calculation shows that spatial structure of the two-photon detection amplitude of the down-converted beams carries important information about conservation or non-conservation of orbital angular momentum in spontaneous parametric down-conversion.
\end{abstract}


\maketitle

\section{\label{sec:intro}Introduction}
The topic whether orbital angular momentum (OAM) is conserved in spontaneous parametric down-conversion (SPDC) has been studied for many years \cite{arlt99, arnaut00, mair01, arnold02, walborn04, terriza03}. Conservation of OAM in the SPDC process has been theoretically attributed to phase matching \cite{arnold02}, transfer of plane-wave spectrum from pump beam to down-converted beams \cite{walborn04}, or to collinear geometry \cite{terriza03}. However, according to quantum mechanics, the conservation of angular momentum (AM) arises from the rotational symmetry of the Hamiltonian describing the studied physical process. Recently, experimental evidence \cite{feng07} has been found which shows that AM non-conservation can occur in SPDC process due to spatial asymmetry of the Hamiltonian. 

It can be theoretically shown \cite{walborn04}, under the paraxial approximation, that OAM is conserved in the SPDC process for thin non-linear media if one assumes that the two-photon detection amplitude (the definition follows soon) in the SPDC process reproduces the pump beam transverse profile. This assumption could be invalid in type-II SPDC process where OAM non-conservation was observed \cite{feng07}, which stimulated our study presented here.

The two-photon detection amplitude in the SPDC process has been extensively explored both theoretically \cite{hong85, rubin94, pittman96, rubin96, monken98} and experimentally \cite{pittman96, monken98, pittman95} without considering the physical variable of OAM. More works have been published on the same topic when non-zero OAM is concerned \cite{arnold02, walborn04, terriza03, ren04, law04, calvo07}. Further inspection shows that all these works implicitly assume rotationally symmetric Hamiltonian for the SPDC process, which is intrinsically related to AM conservation. We investigate the cases where both OAM conservation and OAM non-conservation are taken into account.

We follow Arnaut and Barbosa \cite{arnaut00, barbosa02} in our theoretical treatment. As one will see, there is an essential difference between the notations in \cite{arnaut00, barbosa02} and those used in \cite{arnold02, walborn04, terriza03}.

\vspace{0.1in}
\section{\label{sec:theory1} State of single beam of light with non-zero OAM}

\begin{figure}
\includegraphics[scale=0.35]{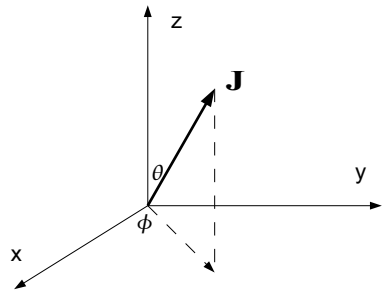}%
\caption{\label{fig1}The vector of OAM {\bf J} in arbitrarily chosen coordinates. $\theta$ and $\phi$ are the polar angle and the azimuthal angle, respectively.}
\end{figure}

In most theoretical works involving the variable of OAM \cite{arnold02,walborn04,terriza03,law04}, coordinates were chosen such that the $z$-axis is along the direction of the OAM carried by the light beam under study for convenience. We consider the general case where the coordinates are arbitrarily chosen (Fig.~\ref{fig1}). 

When the conservation of a vector is concerned, one should always break the vector into components, for example, along $x$, $y$, and $z$-axes. The conservation of the vector means all these components are conserved. If one of the components is not conserved, the vector is said to be non-conserved. The same argument applies to the cases of OAM (non-)conservation in the SPDC process. 

To exam whether the OAM is conserved along the pump propagation direction ($z$-direction) in the SPDC process, one needs to theoretically describe the state of the down-converted light beams that carry OAMs, the $z$-components of which are, say $J_z^i$ and $J_z^s$. 

An elegant approach exploiting orbital Poincar\'{e} spheres to study this general case was provided in \cite{calvo06}, where the $z$-component of OAM carried by photons is, however, not guaranteed to be integer times $\hbar$. We quantize the studied field along $z$-axis. In quantum theory, the eigenstate of the orbital angular momentum operator $\hat{J}_z$ can be found by solving the eigenvalue equation
\begin{equation}
\hat{J}_z|\psi(t)\rangle=\alpha|\psi(t)\rangle, \nonumber
\end{equation}
which leads to a solution for a one-photon field in free space as follows \cite{arnaut00},
\begin{equation}
|\psi^l(t)\rangle=
\int d^3k \ g({\bf k},t)e^{il\phi_{\bf k}}\hat{a}^\dagger({\bf k})|0\rangle,
\label{phi1}
\end{equation}
where $\alpha=l \hbar$ is the $z$-component of the OAM carried by the photon ($l$ is any integer), and $\phi_{\bf k}$ the azimuthal angle of the wave vector ${\bf k}$. The $g({\bf k},t)$ is a function {\em independent} of the azimuthal angle $\phi_{\bf k}$ and, therefore, can be written in the form $g(k_\rho, k_z, t)$, where $k_\rho=\sqrt{{\bf k}^2-k_z^2}$ is the amplitude of the transverse component ${\bf k}_\rho$ of ${\bf k}$. The $\hat{a}^\dagger({\bf k})$ is the photon creation operator. One notes that the freedom of the field polarization is neglected and emphasizes that the $l \hbar$ in Eq. (\ref{phi1}) has a physical meaning (OAM along the $z$-axis) that is essentially {\em different} from what means by the same notation in \cite{arnold02,walborn04,terriza03}, where the $l \hbar$ represents the OAM carried by photons along the axes dictated by the {\it signal}, {\it idler}, or pump central vectors.

The one-photon detection amplitude $\varphi_1^l({\bf r})\equiv\langle0|\hat{E}^{(+)}({\bf r})|\psi^l(t)\rangle$, where $\hat{E}^{(+)}({\bf r})=\int d^3kC_{k}\hat{a}({\bf k})e^{i{\bf k}\cdot{\bf r}}$ [$\hat{a}$({\bf k}) is the annihilation operator, $C_{k}$ is a coefficient that depends on the amplitude of $k=|{\bf k}|$], for the one-photon field in the eigenstate of the operator $\hat{J}_z$ is
\begin{eqnarray}
\varphi_1^l({\bf r})=
\int d^3k \ h(k_\rho,k_z,t)e^{il\phi_{\bf k}}e^{i{\bf k}\cdot{\bf r}}, 
\nonumber
\end{eqnarray}
where $h(k_\rho,k_z,t)=g(k_\rho,k_z,t)C_{k}$. In a plane ($z=z_0$) transverse to the $z$-axis, the one-photon detection amplitude will be 
\begin{equation}
\varphi_1^l(\vec{\rho})=
\int d^2k_\rho \ h(k_\rho,t)e^{il\phi_{{\bf k}_\rho}}e^{i{\bf k}_\rho\cdot\vec{\rho}},
\end{equation}
where $\phi_{{\bf k}_\rho}=\phi_{\bf k}$, and $h(k_\rho,t)=\int dk_z h(k_\rho,k_z,t)e^{ik_z z_0}$. Now we show that $|\varphi_1^l({\bf k}_\rho)|$ is azimuthally symmetric, i.e., it is invariant under the operation of rotation around the $z$-axis (azimuthal rotation) defined as $\hat{R}: \vec{\rho}\rightarrow \vec{\rho}\ '=\hat{R}\vec{\rho}$, where $\rho=\rho'$ and $\phi_{\rho'}=\phi_{\rho}+\delta\phi$.

Under the operation $\hat{R}$, the function $\varphi_1^l(\vec{\rho})$ is transformed into a new one: ${\varphi'}_1^l(\vec{\rho})=\varphi_1^l(\hat{R}^{-1}\vec{\rho})=\int d^2\nu h(\nu)e^{il\phi_{\nu}}e^{i\vec{\nu}\cdot(\hat{R}^{-1}\vec{\rho})}$. Given that $\vec{\nu}\cdot(\hat{R}^{-1}\vec{\rho})=(\hat{R}\vec{\nu})\cdot\vec{\rho}$, ${\varphi'}_1^l(\vec{\rho})=\int d^2\nu h(\nu)e^{il\phi_{\nu}}e^{i(\hat{R}\vec{\nu})\cdot\vec{\rho}}$. With the integral variable $\vec{\mu}\equiv\hat{R}\vec{\nu}$ ($\mu=\nu$ and $\phi_{\mu}=\phi_{\nu}+\delta \phi$), ${\varphi'}_1^l(\vec{\rho})=e^{-il\delta \phi}\int d^2\mu h(\mu)e^{il\phi_{\mu}}e^{i\vec{\mu}\cdot\vec{\rho}}=e^{-il\delta \phi}\varphi_1^l(\vec{\rho})$ [please note that $h$ is independent of the azimuthal angle, i.e., $h(\nu)=h(\mu)$], which means invariant modulus for $\varphi_1^l(\vec{\rho})$ under the rotational operation: $|{\varphi'}_1^l(\vec{\rho})|= |\varphi_1^l(\vec{\rho})|$. 

It can be further shown, in a general case, that the one-photon detection amplitude 
\begin{equation}\label{opda}
\varphi_1^l(\vec{\rho},\vec{\rho}_0)=
\int d^2k_\rho \ h(k_\rho,t)e^{il\phi_{{\bf k}_\rho}}e^{i{\bf k}_\rho\cdot(\vec{\rho}-\vec{\rho}_0)}
\end{equation}
is azimuthally symmetric around the center point $\vec{\rho}_0$ that may be a function of $z_0$. $\varphi_1^l(\vec{\rho},\vec{\rho}_0)$ can be obtained from $\varphi_1^l({\bf r})$ by coordinate transform of ${\bf r}\rightarrow {\bf r}-\vec{\rho}_0(z)$ and that $\vec{\rho}_0(z)$ represents a curve that is comprised of points where the centers of $\varphi_1^l(\vec{\rho},\vec{\rho}_0)$ are located.
\vspace{0.0in}
\section{\label{sec:theory2} State of down-converted twin-beams in SPDC process}
For type-I SPDC process, where the OAM conservation rule holds \cite{mair01}, the state vector of the down-converted light is calculated in \cite{arnaut00} to the first order approximation. Here we consider the general case, in which the OAM conservation may be violated in the SPDC process.

\begin{figure}
\includegraphics[scale=0.3]{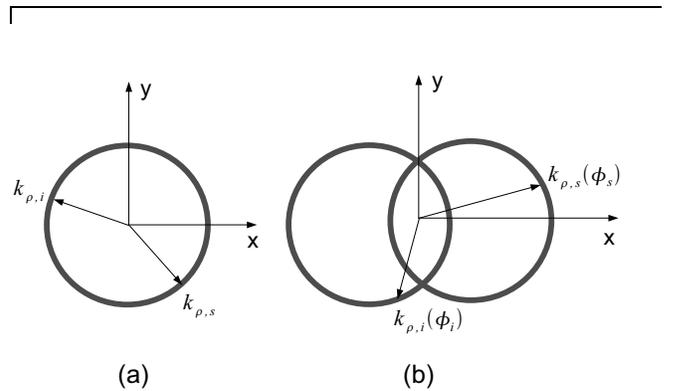}%
\caption{\label{fig2} Catoons showing typical down-conversion patterns of SPDC processes. (a) Down-conversion pattern of the type-I SPDC process, which is azimuthally symmetric. The amplitudes $k_{\rho,s}\ (k_{\rho,i})$ of the transverse components ${\bf k}_{\rho,s}\ ({\bf k}_{\rho,i})$ of the wave vectors are independent of the azimuthal angles $\phi_s\ (\phi_i)$. (b) Down-conversion pattern of the type-II SPDC process, which is azimuthally asymmetric. The amplitudes $k_{\rho,s}\ (k_{\rho,i})$ of the transverse components ${\bf k}_{\rho,s}\ ({\bf k}_{\rho,i})$ of the wave vectors are functions of the azimuthal angles $\phi_s\ (\phi_i)$.}
\end{figure}

Assuming a classical pump beam, two down-converted modes, {\it signal} and {\it idler}, which are initially empty, and linear polarization for all involved light beams, the Hamiltonian describing the non-linear process of SPDC in the interaction picture is \cite{arnaut00}
\begin{widetext}
	\begin{eqnarray}
	\label{eq:ham}
\hat{H}_I &=& \int d^3k_sd^3k_il^{(*)}_E(\omega_{{\bf k}_s})l^{(*)}_E(\omega_{{\bf k}_i}){\hat{a}}^\dagger({\bf k}_s){\hat{a}}^\dagger({\bf k}_i) e^{i(\omega_{{\bf k}_s}+\omega_{{\bf k}_i})t} \nonumber \\
& & \chi_{ijn}({\bf e}_{{\bf k}_s})_j^* ({\bf e}_{{\bf k}_i})_n^* \int_{V_I}d^3r{\bf E}_i({\bf r},t)e^{-i({\bf k}_s+{\bf k}_i)\cdot {\bf r}} + H.c.\, ,
	\end{eqnarray}
\end{widetext}
where $V_I$ is the non-linear interaction volume, $l^{(*)}_E(\omega_{\bf k}) = -i\sqrt{\hbar \omega_{\bf k}/2 \varepsilon({\bf k})}$, ${\bf e}_{{\bf k}_{s,i}}$ are unit vectors representing the linear polarizations of the down-converted modes, and ${\bf E}({\bf r},t)$ is the electrical field associated with the pump beam. Subscripts {\it s} and {\it i} indicate {\it signal} and {\it idler} and are completely interchangeable for degenerate cases. A Laguerre-Gaussian pump beam is assumed propagating along $\hat{z}$ with the principal component polarized along $\hat{x}$ in cylindrical coordinates \cite{arnaut00}
\begin{widetext}
	\begin{eqnarray}
	\label{eq:pump}
{\bf E}(\rho,\phi,z;t) \equiv \psi_{lp}({\bf r})e^{i(k_pz-\omega_pt)}\hat{x} &=& \frac{A_{lp}}{\sqrt{1+(z^2/z^2_R)}} \left[ \frac{\sqrt{2}\rho}{w(z)}\right]^lL^l_p\left[\frac{2\rho^2}{w^2(z)}\right]exp\left[i\frac{k_p\rho^2}{2q(z)}\right]\nonumber\\ & & e^{il\tan^{-1}(y/x)} exp\left[-i(2p+l+1)\tan^{-1}(z/z_R)\right]e^{i(k_pz-\omega_pt)}\hat{x}\, ,
	\end{eqnarray}
\end{widetext}
where $z_R$ is the Rayleigh length, $w(z)=w_0\sqrt{1+z^2/z^2_R}$, $w_0$ is the beam radius at the waist $z=0$. $q(z)=z-iz_R$, and $\rho=\sqrt{x^2+y^2}$. To the first order approximation, the state vector of the down-converted light beams reads \cite{arnaut00}
\begin{widetext}
	\begin{eqnarray}
	\label{eq:onewave}
|\psi(t)\rangle = |0\rangle + \int d^3k_s \int d^3k_i A_{{\bf k}_s,{\bf k}_i} l^{(*)}_E(\omega_{{\bf k}_s})l^{(*)}_E(\omega_{{\bf k}_i}) T(\Delta \omega)\tilde{\psi}_{lp}(\Delta {\bf k}){\hat{a}}^\dagger({\bf k}_s){\hat{a}}^\dagger({\bf k}_i)|0\rangle\, ,
	\end{eqnarray}
\end{widetext}
where\\ $A_{{\bf k}_s,{\bf k}_i} = A_{lp}\chi_{1jm}[({\bf e}_{{\bf k}_s})^*_j({\bf e}_{{\bf k}_i})^*_m+({\bf e}_{{\bf k}_i,i})^*_j({\bf e}_{{\bf k}_s,s})^*_m]$, $T(\Delta \omega)=exp[i\Delta \omega(t-t_{int}/2)]\sin(\Delta \omega t_{int}/2)/(\Delta \omega/2)$ is the time window function defining the $\Delta \omega$ range given the interaction time $t_{int}$, $\Delta \omega=\omega_{{\bf k}_s}+\omega_{{\bf k}_i}-\omega_{p}$, $\Delta {\bf k}={\bf k}_s+{\bf k}_i-{\bf k}_p$, and $\tilde{\psi}_{lp}(\Delta {\bf k})=\int_{V_I}d^3r{\psi}_{lp}({\bf r})exp(-i\Delta {\bf k}\cdot{\bf r})$.
Under the usual conditions that the non-linear medium is centered on the $z$ axis, the average radius of the beam is small compared to the transverse section of the non-linear medium and that the crystal length is smaller than the Rayleigh range $z_R$ of the pump beam, one has \cite{arnaut00}
\newcommand{\bin}[2]{
\left( 
      \begin{array}
             {@{}c@{}}
             #1 \\ #2
      \end{array} 
\right)
}
\begin{widetext}
	\begin{eqnarray}
	\label{eq:geo}
\tilde{\psi}_{lp}(\Delta {\bf k}) = B_{lp}W(\Delta k_z)\phi_{lp}(k_{\rho,s},k_{\rho,i},\phi_s,\phi_i)\sum\limits_{n=0}^l\bin{l}{n}(k_{\rho,s})^n(k_{\rho,i})^{l-n}e^{in\phi_s}e^{i(l-n)\phi_i} \, ,
	\end{eqnarray}
where $B_{lp}=(z_R\pi/k^{l+1}_p)(\sqrt{2}z_R/w_0)^lexp[-\frac{\pi}{2}i(1-l-p)]2^{p-l+1}$, $z_0$ is the position of the center of the non-linear medium, $W(\Delta k_z)=exp(-i\Delta k_z z_0)[\sin(\Delta k_zl/2)/(\Delta k_z/2)]$, $l$ is the medium length, $k_{\rho,s}$ and $k_{\rho,i}$ are amplitudes of the transverse components ${\bf k}_{\rho,s}$ and ${\bf k}_{\rho,i}$ of the wave vectors, and 
	\begin{eqnarray}
	\label{eq:core}
\phi_{lp}(k_{\rho,s},k_{\rho,i},\phi_s,\phi_i) = L_p^l\left(\frac{z_R}{k_p}\rho_k^2\right)exp\left(-\frac{z_R}{2k_p}\rho_k^2\right)exp\left(-i\frac{z_0}{2k_p}\rho_k^2\right) \, ,
	\end{eqnarray}
where 
 	\begin{eqnarray}
	\label{eq:rho}
\rho_k\equiv|\Delta k_x +\Delta k_y|& = &\sqrt{k_{\rho,s}^2+k_{\rho,i}^2+2k_{\rho,s}k_{\rho,i}\cos(\phi_s-\phi_i)}  \, ,
	\end{eqnarray}
\end{widetext}
where $\phi_s$ and $\phi_i$ are the azimuthal angles of the transverse wave vectors ${\bf k}_{\rho,s}$ and ${\bf k}_{\rho,i}$, respectively. It is well known that, in the type-I case, the down-conversion is a single-ring pattern, which has azimuthally spatial symmetry [Fig. \ref{fig2}(a)]. Accordingly, the amplitudes $k_{\rho,s}$ and $k_{\rho,i}$ are independent of $\phi_s$ and $\phi_i$. Then $\rho_k$ is a function of $\phi_s-\phi_i$ and  $\phi_{lp}(k_{\rho,s},k_{\rho,i},\phi_s,\phi_i) $ can be expanded in terms of $\phi_s-\phi_i$ as given by Eq. (15) in \cite{arnaut00}. However, the azimuthal symmetry may not always exist in the SPDC process, an example of which is the type-II SPDC [Fig. \ref{fig2}(b)]. In this case, $k_{\rho,s}$ and $k_{\rho,i}$ are dependent of $\phi_s$ and $\phi_i$, respectively. As a consequence, Eq. (15) of \cite{arnaut00} should be replaced with a more general expansion as follows:
\begin{widetext}
 	\begin{eqnarray}
	\label{eq:phi}
\phi_{lp}(k_{\rho,s},k_{\rho,i},\phi_s,\phi_i)(k_{\rho,s})^n(k_{\rho,i})^{l-n} = \sum\limits_{m_s,m_i=-\infty}^{\infty} G^{m_s,m_i,n}_{lp}e^{-i(m_s\phi_s-m_i\phi_i)} \, ,
	\end{eqnarray}
where $G^{m_s,m_i,n}_{lp}$ is a coefficient that does not depend on $\phi_s$ or $\phi_i$. Substituting Eq. (\ref{eq:geo}) and Eq. (\ref{eq:phi}) into Eq.~(\ref{eq:onewave}), one arrives at
	\begin{eqnarray}
	\label{eq:newstate}
|\psi(t)\rangle &=& \int d^3k_s\int d^3k_i B_{lp}A_{{\bf k}_s,{\bf k}_i} l^{(*)}_E(\omega_{{\bf k}_s})l^{(*)}_E(\omega_{{\bf k}_i}) T(\Delta \omega)W(\Delta k_z)\nonumber \\ 
 & & \sum\limits_{n=0}^l\ \sum\limits_{m_s,m_i=-\infty}^{\infty}G^{m_s,m_i,n}_{lp}\bin{l}{n}e^{i(n-m_s)\phi_s}e^{i(l-n+m_i)\phi_i}\hat{a}^\dagger({\bf k}_s)\hat{a}^\dagger({\bf k}_i)|0\rangle\, ,
	\end{eqnarray}
\end{widetext}
where the vacuum term is dropped. The physical meaning of the $n-m_s$ and $l-n+m_i$ here is that the $z$-component of the OAM carried by each {\it signal} or {\it idler} photon is $(n-m_s)\hbar$ or $(l-n+m_i)\hbar$ \cite{arnaut00}. So, the total OAM carried by a pair of down-converted photons is $l_{s+i}\hbar=(n-m_s)\hbar+(l-n+m_i)\hbar=(l-m_s+m_i)\hbar$. It is apparent that the OAM is conserved along the $z$-axis in the SPDC process if and only if $m_s=m_i$. So the two-photon detection amplitude of the down-converted beams carries information about whether the OAM is conserved in the SPDC process.

Similar to the definition of the one-photon detection amplitude, the two-photon detection amplitude is defined as $\varphi_2({\bf r}_s,{\bf r}_i)\equiv\langle0|\hat{E}_s^{(+)}({\bf r}_s)\hat{E}_i^{(+)}({\bf r}_i)|\psi(t)\rangle$, which can be calculated for the case of the SPDC process with Eq. (\ref{eq:newstate}).
\begin{widetext}
	\begin{eqnarray}
\varphi_2({\bf r}_s,{\bf r}_i)=\int d^3k_sd^3k_i \sum\limits_{n=0}^l\ \sum\limits_{m_s,m_i=-\infty}^{\infty}F^{m_s,m_i,n}_{lp}({\bf k}_s,{\bf k}_i)e^{i(n-m_s)\phi_s+i(l-n+m_i)\phi_i}e^{i[{\bf k}_s\cdot({\bf r}_s-\vec{\rho}_{0,s})+{\bf k}_i\cdot({\bf r}_i-\vec{\rho}_{0,i})]}\nonumber\, ,
	\end{eqnarray}
where
 	\begin{eqnarray}
	\label{eq:prob}
F^{m_s,m_i,n}_{lp}({\bf k}_s,{\bf k}_i) = C_{k_s}C_{k_i}A_{{\bf k}_s,{\bf k}_i} l^{(*)}_E(\omega_{{\bf k}_s})l^{(*)}_E(\omega_{{\bf k}_i}) B_{lp} T(\Delta \omega)W(\Delta k_z)G^{m_s,m_i,n}_{lp}\bin{l}{n}\nonumber\, ,
	\end{eqnarray}
and $\vec{\rho}_{0,s}, \vec{\rho}_{0,i}$ are the transverse coordinates of the centers of the transverse profiles of the {\it signal} and {\it idler} modes. The transverse profile of $\varphi_2({\bf r}_s,{\bf r}_i)$ can be obtained by setting $z_s=z_{0,s}$, $z_i=z_{0,i}$, which leads to 
	\begin{eqnarray}
	\label{tpda}
\varphi_2(\vec{\rho}_s,\vec{\rho}_i)&=&\int d^2k_{\rho,s}d^2k_{\rho,i} \sum\limits_{n=0}^l\ \sum\limits_{m_s,m_i=-\infty}^{\infty}H^{m_s,m_i,n}_{lp}({\bf k}_{\rho,s},{\bf k}_{\rho,i})\nonumber \\
& & e^{i(n-m_s)\phi_s+i(l-n+m_i)\phi_i}e^{i[{\bf k}_{\rho,s}\cdot(\vec{\rho}_s-\vec{\rho}_{0,s})+{\bf k}_{\rho,i}\cdot(\vec{\rho}_i-\vec{\rho}_{0,i})]}\, ,
	\end{eqnarray}
where $H^{m_s,m_i,n}_{lp}({\bf k}_{\rho,s},{\bf k}_{\rho,i})= \int dk_{z,s}dk_{z,i}F^{m_s,m_i,n}_{lp}({\bf k}_s,{\bf k}_i)e^{i(k_{z,s}z_{0,s}+k_{z,i}z_{0,i})}$.
\end{widetext}

In the degenerate case, we have approximately ${\bf k}_{\rho,s}+{\bf k}_{\rho,i}\approx0$ ($k_{\rho,s}\approx k_{\rho,i}$ and $\phi_s\approx\phi_i+\pi$), which can be justified by plugging regular experimental parameters (Rayleigh range $Z_R\gg$1cm and $l_p\approx5$ for pump beam) into Eq. (13) and Eq. (16) of \cite{barbosa02}. Applying these approximations in the form of $\delta({\bf k}_{\rho,s}+{\bf k}_{\rho,i})$ to Eq. (\ref{tpda}), one obtains
\begin{widetext}
	\begin{eqnarray}
	\label{tpda1}
\varphi_2(\vec{\rho}_s,\vec{\rho}_i)=e^{-i(l-n+m_i)\pi}\int d^2k_{\rho,s} \sum\limits_{n=0}^l\ \sum\limits_{m_s,m_i=-\infty}^{\infty}H^{m_s,m_i,n}_{lp}({\bf k}_{\rho,s},-{\bf k}_{\rho,s})e^{i(l-m_s+m_i)\phi_s}e^{i{\bf k}_{\rho,s}\cdot(\vec{\rho}_s-\vec{\rho}_i-\vec{\rho}_{0,s}+\vec{\rho}_{0,i})}\, .
	\end{eqnarray}
\end{widetext}

Now suppose that one fixed point detector is located at $\vec{\rho}_{0,i}$ in the {\it idler} beam while another detector is located on the {\it signal} side to measure the two-photon detection amplitude of the down-converted beams. After simple mathematical manipulation, one arrives at

\begin{widetext}
	\begin{eqnarray}
	\label{tpda2}
\varphi_2(\vec{\rho}_s,\vec{\rho}_{0,i})=e^{-i(l-n+m_i)\pi}\sum\limits_{m=-\infty}^{\infty}\int d^2k_{\rho,s} G^m_p({\bf k}_{\rho,s})e^{im\phi_s}e^{i{\bf k}_{\rho,s}\cdot(\vec{\rho}_s-\vec{\rho}_{0,s})}\, ,
	\end{eqnarray}
where $G^m_p({\bf k}_{\rho,s})=\sum\limits_{n=0}^l\ \sum\limits_{m_s=-\infty}^{\infty}H^{m_s,m-l+m_s,n}_{lp}({\bf k}_{\rho,s},-{\bf k}_{\rho,s})$, and $m=l-m_s+m_i$. $m\hbar$ represents the total OAM along $z$-axis carried by a pair of down-converted photons.
\end{widetext}
If the dependence of the term $A_{{\bf k}_s,{\bf k}_i}$ and the refractive indices on the wave vectors is negligible, then $G^m_p({\bf k}_{\rho,s})$ is independent of the azimuthal angle $\phi_s$ of the wave vector $k_{\rho,s}$. According to Eq. (\ref{opda}), the two-photon detection amplitude of the down-converted beams can be further written as
	\begin{eqnarray}
	\label{tpda2}
\varphi_2(\vec{\rho}_s,\vec{\rho}_{0,i})=e^{-i(l-n+m_i)\pi}\sum\limits_{m=-\infty}^{\infty}\varphi_1^m(\vec{\rho}_s,\vec{\rho}_{0,s})\, ,
	\end{eqnarray}
where $\varphi_1^m(\vec{\rho}_s,\vec{\rho}_{0,s})=\int d^2k_{\rho} G^m_p(k_{\rho})e^{im\phi}e^{i{\bf k}_{\rho}\cdot(\vec{\rho}_s-\vec{\rho}_{0,s})}$, $\phi$ is the azimuthal angle of ${\bf k}_{\rho}$. Neglecting the overall phase factor, Eq. (\ref{tpda2}) shows that the transverse profile of the two-photon detection amplitude of the down-converted beams, in general, equals the sum of one-photon detection amplitudes of light beams that carry OAMs [$m\hbar=(l-m_s+m_i)\hbar$ per photon along $z$-axis]. As pointed out in previous context, whether OAM is conserved can be justified by the value of $m=l-m_s+m_i$. So, Eq. (\ref{tpda2}) reveals whether the OAM is conserved or not in the SPDC process. This is the major discovery in our theoretical analysis presented here. 

\begin{figure}
\vspace{0.2in}
\centerline{
\includegraphics[scale=0.3]{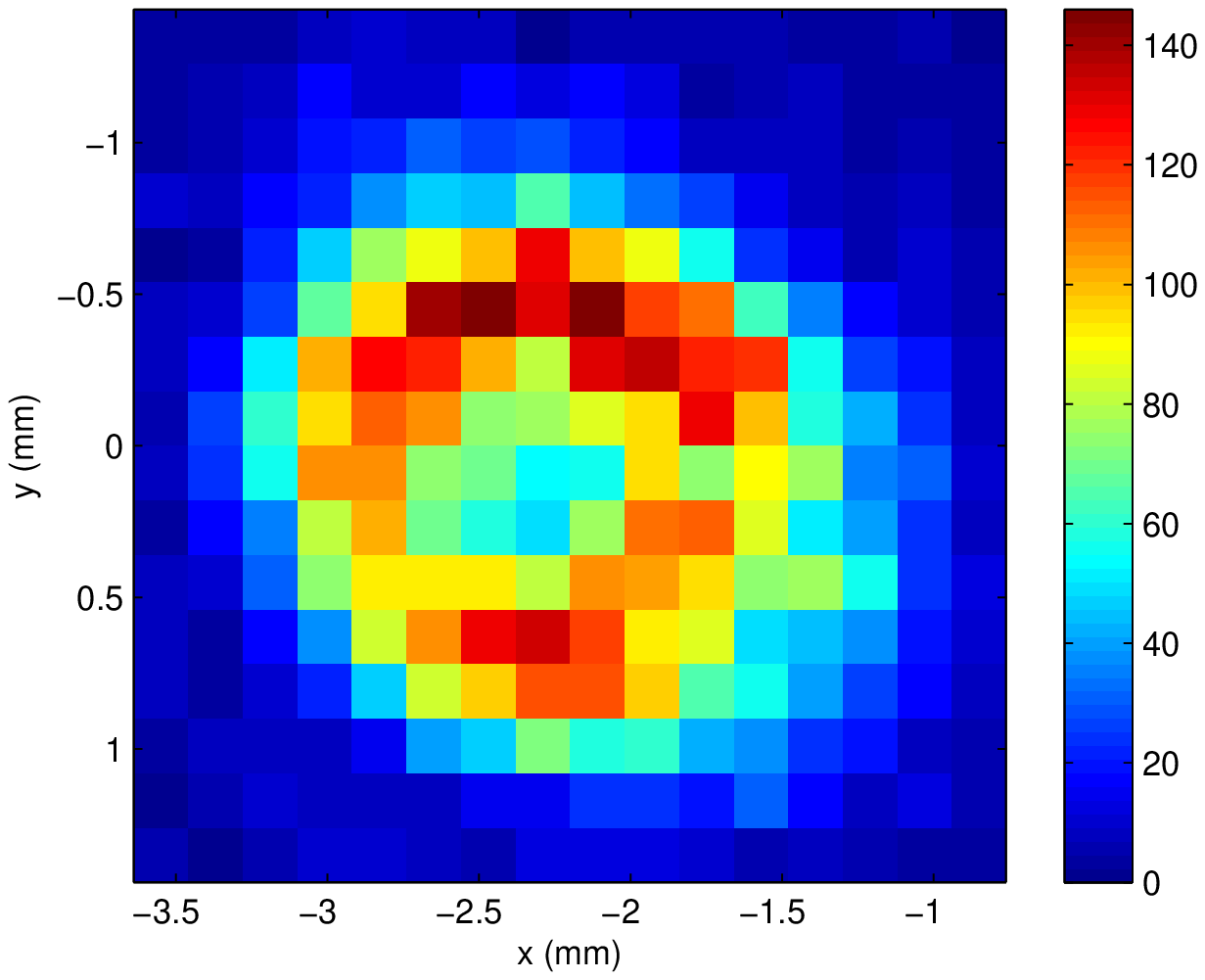}
\hspace{0.1in}
\includegraphics[scale=0.3]{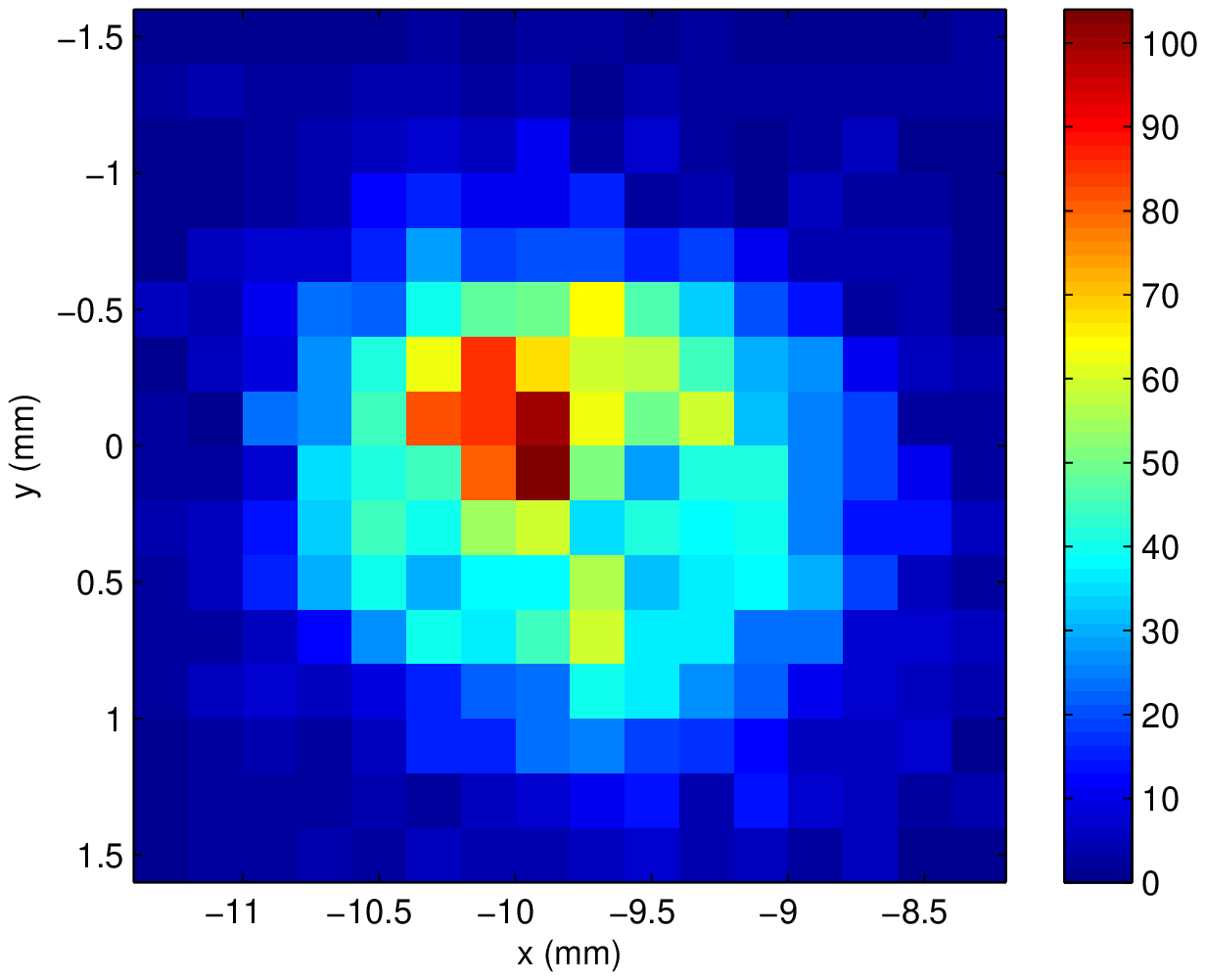}
}
\caption{\label{exp}(color online) Transformation of the transverse profile of the two-photon detection amplitude of the down-converted beams in the SPDC process with holographic masks. (Left) The transverse profile of the two-photon detection amplitude of the down-converted beams in type-I SPDC process, where the OAM is conserved \cite{mair01}, with pump beam carrying non-zero OAM ($l\hbar$ per photon along $z$-axis, $l=2$). (Right) Transformed transverse profile, which is two-dimensionally Gaussian according our data-fitting results, of the two-photon detection amplitude with a holographic mask ($l=-2$). }
\end{figure}

The simplest case is that the OAM is conserved in the SPDC process, i.e., $m_s=m_i$. Then $\varphi_2(\vec{\rho}_s,\vec{\rho}_{0,i})=e^{-i(l-n+m_i)\pi}\varphi_1^l(\vec{\rho}_s,\vec{\rho}_{0,s})$. This shows that $|\varphi_2(\vec{\rho}_s,\vec{\rho}_{0,i})|$ possesses the same azimuthal symmetry as $|\varphi_1^l(\vec{\rho}_s,\vec{\rho}_{0,s})|$, which is, according to Section \ref{sec:theory1}, azimuthally symmetric around point $\vec{\rho}_{0,s}$ . In the case of small polar angle, a light beam carrying OAM ($m\hbar$ per photon along $z$-axis) approximately has a Laugerre-Gaussian (LG) profile, which can be transformed into a Gaussian one with an appropriately chosen holographic mask (-$m$)\cite{mair01}. So does $\varphi_2(\vec{\rho}_s,\vec{\rho}_{0,i})$ with a holographic mask (-$l$) when the OAM is conserved, which agrees with experimental observations (Fig. \ref{exp}).

Now we consider the case of OAM non-conservation in the SPDC process. Assuming that $m_s\ne m_i$, but $m=l-m_s+m_i$ has a fixed value, say $m_a$. Then only one term in Eq. (\ref{tpda2}) is non-zero: $\varphi_2(\vec{\rho}_s,\vec{\rho}_{0,i})=e^{-i(l-n+m_i)\pi}\varphi_1^{m_a}(\vec{\rho}_s,\vec{\rho}_{0,s})$, where $m_a\ne l$. In this case, the two-photon detection amplitude still has a transverse profile that possesses azimuthal symmetry around point $\vec{\rho}_{0,s}$, which means that azimuthal symmetry possessed by the two-photon detection amplitude is only a necessary condition for the OAM to be conserved in the SPDC process. Nevertheless, $\varphi_2(\vec{\rho}_s,\vec{\rho}_{0,i})$ {\em cannot} be transformed into a Gaussian profile with a -$l$ holographic mask . Instead, one needs a different mask (-$m_a$) to do the transformation. We call this case as {\it type-A} OAM non-conservation.

More generally, $m=l-m_s+m_i$ may not have a fixed value. Then $\varphi_2(\vec{\rho}_s,\vec{\rho}_{0,i})$ is a mixture of many terms: $e^{-i(l-n+m_i)\pi}\sum\limits_m\varphi_1^{m}(\vec{\rho}_s,\vec{\rho}_{0,s})$, which does not have azimuthal symmetry around $\vec{\rho}_{0,s}$. The transverse profile of the two-photon detection amplitude is more complicated that the type-A case and can never be transformed into a Gaussian one with any holographic mask (-$n$, $n$ is any integer). We name this case as {\it type-B} OAM non-conservation.

\section{\label{sec:conclusion} Conclusion}

We theoretically show that the two-photon detection amplitude of the down-converted beams in the SPDC process carries information about whether the OAM is conserved or how the conservation is violated in the non-conservation cases. We find that azimuthal symmetry in transverse profile possessed by the two-photon detection amplitude is a necessary condition for the OAM to be conserved in the SPDC process. In other words, azimuthal asymmetry of the two-photon detection amplitude is a sufficient condition for OAM non-conservation. With the help of appropriately chosen holographic masks, one can tell whether the OAM is conserved with certainty when the polar angle is small. If the OAM conservation is violated, one can also use holographic masks to do analysis to find out which type of OAM non-conservation it belongs to. This work was supported in part by the Quantum Imaging MURI funded through the U.S. Army Research Office. 

\end{document}